\documentclass[12pt,russian]{article}
 \usepackage{helvet}
 \usepackage[T1]{fontenc}
 \usepackage[cp1251]{inputenc}
 \usepackage{geometry}
 \usepackage{graphicx}
 \usepackage{setspace}
 \usepackage{babel}
 \usepackage{amssymb}

 \renewcommand{\vec}[1]{\mbox{\boldmath $#1$}}

 \def\rot{\mathop{\rm rot}\nolimits}

 \def\gsim{\lower.4ex\hbox{$\;\buildrel >\over{\scriptstyle\sim}\;$}}
 \def\lsim{\lower.4ex\hbox{$\;\buildrel <\over{\scriptstyle\sim}\;$}}
 \def\newpage{\vfill\eject}
 \def\bl{\par\vskip 12pt\noindent}
 \def\bll{\par\vskip 24pt\noindent}

 \def\beg{\begin{eqnarray}}
 \def\ende{\end{eqnarray}}

 \def\aa{Astron. Astrophys.}
 \def\apj{Astrophys.~J.}

 \def\nat{Nature}
 \def\sp{Sol. Phys.}

\begin{document}

\begin{center}
{\bf FLUCTUATIONS IN THE ALPHA-EFFECT \\ AND GRAND SOLAR MINIMA}
\end{center}

\bll

\centerline{S.\,V.~Olemskoy$^1$, A.\,R.~Choudhuri$^2$,
and L.\,L.~Kitchatinov$^{1,3}$\footnote{E-mail: kit@iszf.irk.ru} }

\bl

\begin{center}
$^1${\it Institute of Solar-Terrestrial Physics, Lermontov Str. 126A, Irkutsk 664033, Russia} \\
$^2${\it Department of Physics, Indian Institute of Science, Bangalore-560012} \\
$^3${\it Pulkovo Astronomical Observatory, St. Petersburg 176140, Russia}
\end{center}

\bll
\centerline{\bf ABSTRACT}
\bll

%\hspace{0.8 truecm}
\parbox{14.4 truecm}{
Parameters of a special kind of $\alpha$-effect known in
dynamo theory as the Babcock–Leighton mechanism are estimated using the
data of sunspot catalogs. The estimates evidence the presence of the Babcock–Leighton
$\alpha$-effect on the Sun. Fluctuations of the $\alpha$-effect are also estimated.
The fluctuation amplitude appreciably exceeds the mean value, and the characteristic
time for the fluctuations is comparable to the period of the solar rotation.
Fluctuations with the parameters found are included in a numerical model
for the solar dynamo. Computations show irregular changes in the amplitudes
of the magnetic cycles on time scales of centuries and millennia. The calculated
statistical characteristics of the grand solar minima and maxima agree with the
data on solar activity over the Holocene.
 }

\bll

DOI: 10.1134/S1063772913050065

\newpage

\reversemarginpar

\setlength{\baselineskip}{0.8 truecm}

%%%%%%%%%%%%%%%%%%%%%%%%%%%%%%%%%%%%%%%%%%%%%%%%%%%%%%%%%%%%%%%%%%%
 \centerline{1.~INTRODUCTION}
 \bl
%%%%%%%%%%%%%%%%%%%%%%%%%%%%%%%%%%%%%%%%%%%%%%%%%%%%%%%%%%%%%%%%%%%
The purpose of the present work is twofold. First,
we attempt to estimate the parameters of the solar dynamo $\alpha$-effect
from sunspot data. We then use the found parameters in a numerical model of the
solar dynamo to reproduce the grand minima of solar activity.

Dynamo theory explains solar magnetic activity
in terms of two main effects: toroidal field production from the global
poloidal field by differential rotation (the $\Omega$-effect) and the inverse
transformation of the toroidal field into the poloidal field by cyclonic flows
(the $\alpha$-effect). The resulting mechanism for the field generation
is called the $\alpha\Omega$-dynamo (see, for example, [\ref{KR84}]).

Differential solar rotation varies slightly with
time. There are only torsional oscillations with the
11-year period of the solar cycle and an amplitude
of several m/s [\ref{LH82},\ref{Vea02}]. Therefore, the $\Omega$-effect is fairly
regular. The presence of this effect on the sun is evidenced by the observed dependence
of the cycle amplitudes on the strengths of the poloidal (polar)
field at the preceding solar minima [\ref{MT00}-\ref{JCC07}].
At the same time, there is no correlation between the
amplitudes of the solar cycles and the poloidal fields
at {\it following} solar minima. The reason is
believed to be a significant randomness in the $\alpha$-effect [\ref{C11}].
The $\alpha$-effect on the
sun and other convective stars is related to small-scale
cyclonic motions. Various versions of this
effect differ only in the origin of the corresponding
small-scale flows. These can be either convective
flows [\ref{P55}], or motions due to magnetic buoyancy [\ref{B61},\ref{L69}].
The $\alpha$-effect associated with the buoyant rise of toroidal
fields to the solar surface was named the Babcock–Leighton mechanism.

The Babcock–Leighton mechanism deserves
special attention for two reasons. First, in contrast
to other types of $\alpha$-effect, this
mechanism does not suffer from catastrophic quenching
due to the conservation of magnetic helicity
[\ref{KO11a},\ref{KO11b}]. Therefore, this type of $\alpha$-effect may
dominate on the Sun. Second, parameters of
the Babcock–Leighton mechanism can be estimated
from the sunspot data [\ref{E04},\ref{DEea10}]. Recent
estimations for three solar cycles [\ref{KO11c}] evidence operation of
this mechanism on the Sun. The analysis of this paper relies on longer data
series. We estimate both the parameters of the Babcock–Leigh\-ton mechanism for individual
solar cycles and fluctuations of these parameters with time.

Fluctuations of the $\alpha$-effect are important for the
theory of the solar dynamo. The grand minima of solar activity can be interpreted in terms of these fluctuations. Obser\-va\-ti\-ons reveal the alternations of "usual"\ 11-year cycles with grand solar minima, the Maunder minimum being the best known example. Grand minima have also been detected on
solar-type stars [\ref{SB92}]. The most popular - though not the only known - theory of
grand minima explains them by fluctuations of dynamo parameters, mainly by the fluctuations of the $\alpha$-effect (see, for example, [\ref{C92} - \ref{USM09}]). The main difficulty in this explanation is that too strong fluctuations comparable or even exceeding the mean values are required. It is difficult to imagine such fluctuations for the $\alpha$-effect due to convective turbulence. There are likely several tens of global convective cells simultaneously on the sun [\ref{MBT06}]. The relative amplitude of the fluctuations is inversely proportional to the square root of the cells number, and must be below unity. We expect, however, the Babcock–Leighton mechanism to be free from this difficulty. This mechanism is related to solar active regions. Even at solar maxima, there are only a few active regions simultaneously on the Sun. Therefore, we expect fluctuations in the Babcock–Leighton mechanism to be comparable by the order of magnitude to its mean parameters.

In the next section, we estimate the parameters of
the Babcock–Leighton mechanism and their fluctuations
using sunspot data. Section 3 describes the dynamo
model with a non-local $\alpha$-effect corresponding
to the Babcock–Leighton mechanism. The model allows for fluctuations in the $\alpha$-effect whose amplitude is specified in accordance with our obser\-va\-ti\-ons-based estimates. We discuss results of long-term computations covering about one thousand magnetic cycles. Section 4
formulates our main conclusions.
%%%%%%%%%%%%%%%%%%%%%%%%%%%%%%%%%%%%%%%%%%%%%%%%%%%%%%%%%%%%%%%%%%%%
 \bll
 %\newpage
 \begin{center}
 2. ESTIMATING THE BABCOCK-LEIGHTON MECHANISM \\ FROM SUNSPOT DATA
 \end{center}
%%%%%%%%%%%%%%%%%%%%%%%%%%%%%%%%%%%%%%%%%%%%%%%%%%%%%%%%%%%%%%%%%%%%
 \bl
 \centerline{\sl 2.1. Method}
 \bl
%%%%%%%%%%%%%%%%%%%%%%%%%%%%%%%%%%%%%%%%%%%%%%%%%%%%%%%%%%%%%%%%%%%%
The Babcock–Leighton mechanism is related to
Joy’s law for solar active regions. The law states that
the leading spots (in the rotational motion) of bipolar
groups are, on average, located closer to the equator
than the trailing spots. Thus, on average, the line
connecting the centers of opposite polarities displays
a (positive) tilt to the solar lines of latitude (Fig.~1).
The mean tilt angle $\alpha$ increases with latitude, as must
be the case if Joy’s law is due to the Coriolis force
influence on the emerging magnetic loops [\ref{WS89}].

\begin{figure}[htb]
 \centerline{
 \includegraphics[width=10cm]{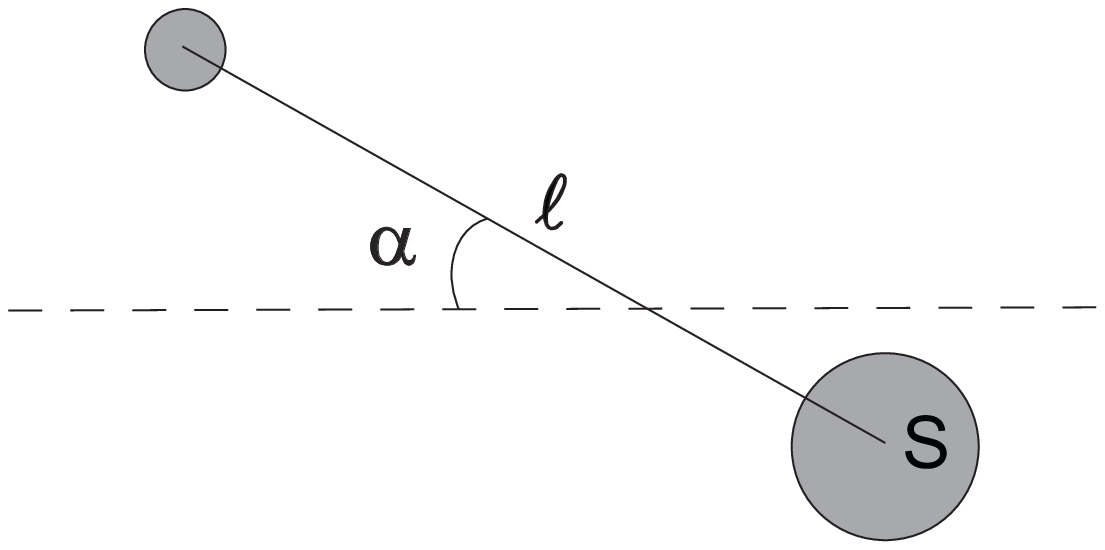}}
 \begin{description}
 \item{\small Fig.~1. Illustration of Joy's law and parameters of Eq.(1).
    The tilt angle $\alpha$) is on average positive, $l$ is the distance
    between the centers of opposite polarities and $S$ is the area of the
    largest spot of the group. Northern hemisphere, the pole is upward.
    }
 \end{description}
\end{figure}

Due to the finite tilt $\alpha$, the magnetic fields of
active regions possess a poloidal compo\-n\-ent. Upon
the decay of active regions and subsequent turbulent
diffusion, this component contributes
to the global poloidal field of the sun. We
assume that the total contribution of active regions to the
poloidal field made over a certain interval of a solar cycle can be estimated as
\begin{equation}
    B = \sum\limits_{i}^{} S_i \ell_i \sin\alpha_i ,
    \label{1}
\end{equation}
where $S_i$ is the area of the largest spot of the group,
$\ell_i$ is the distance between the centers of opposite polarities,
and $\alpha_i$ is the tilt angle of the group axis to
the East–West direction (see Fig. 1). The summed
quantities are taken at the instant of maximum development
of the sunspot group, and the contribution of
each group to the sum (\ref{1}) is taken once.

The estimation assumes that the contribution to the poloidal
field is proportional to the magnetic flux of the active
region. Since the magnetic fields of mature
sunspots vary within a comparatively narrow interval
from about $2.5$ to $3.5$~kG [\ref{O85}], we can assume the magnetic
flux to be proportional to the area of the largest spot.
Typical distances between the spots of a group
are small compared to the solar radius. Turbulent diffusion will annihilate opposite polarities as the group decays. Only a minor part
of the magnetic flux of an active region
contributes to the global poloidal field. Babcock
estimated this part by about 1\% [\ref{B61}]. We
expect this contribution to increase with $\ell$. Since $\ell$
is small compared to $R_\odot$, we assume linear dependence on $\ell$ in the
sum (\ref{1}). Finally, the poloidal field components of active regions are
proportional to $\sin\alpha$.

When the summation in (\ref{1}) is made over the entire
solar cycle, we designate the resulting $B$ value as
$B_\mathrm{cyc}$. Some evidence has been found earlier [\ref{KO11c}]
that the relation between $B_\mathrm{cyc}$ and the poloidal field at the
following activity minima is close to linear depen\-den\-ce. This
indicates in favour of the action of the Babcock–Leighton mechanism on the Sun. We are also interested
in fluctuations in this mechanism, and so
calculate the sums (\ref{1}) for shorter time intervals. The
lifetimes of most sunspots do not exceed one solar rotation.
Therefore, the characteristic time of fluctuations
in the Babcock–Leighton mechanism should not be longer than one solar rotation.
However, it does not seem reasonable to calculate $B$ of (\ref{1})
for times shorter than one Carrington
rotation. We designate the $B$ values for individual
solar rotations as $B_\mathrm{Car}$. These values show significant
fluctuations. Of course, the sum of all the $B_\mathrm{Car}$ values for the individual solar
cycles equal $B_\mathrm{cyc}$. The running average $\langle B_\mathrm{Car}\rangle$ over
13 rotations (approximately one year) varies smoothly
with time, and shows no significant fluctuations. The
relative deviation from this running mean for individual rotations is
\begin{equation}
    B^{'}_\mathrm{Car} = \frac{B_\mathrm{Car}}{\langle B_\mathrm{Car}\rangle }-1 ,
    \label{2}
\end{equation}
while the relative amplitude of the fluctuations is
\begin{equation}
    \sigma_{B_\mathrm{Car}} = \sqrt{\frac{\sum( B^{'}_\mathrm{Car})^2}{N}},
    \label{3}
\end{equation}
where the summation is made over all $N$ rotations
presented in the sunspot catalog.

%%%%%%%%%%%%%%%%%%%%%%%%%%%%%%%%%%%%%%%%%%%%%%%%%%%%%%%%%%%%%%%%%%%%
 \bll
 \centerline{\sl 2.2. Data}
 \bl
%%%%%%%%%%%%%%%%%%%%%%%%%%%%%%%%%%%%%%%%%%%%%%%%%%%%%%%%%%%%%%%%%%%%
Values of $B_\mathrm{cyc}$ were calculated earlier [\ref{KO11c}] for solar
cycles 19–21 using the Catalog of Solar Activity
(CSA) of the Pulkovo Observatory [\ref{Nea08}]
(http://www.gao.spb.ru/database/csa/ groups\_e.html). It
was found that the relationship between the $B_\mathrm{cyc}$
for individual solar cycles and the amplitude
of the poloidal field at the following solar minima is
close to linear dependence. Here, we use the CSA data to estimate
the fluctuations (\ref{2}) and (\ref{3}).
Longer series of data are provided by the Kodaikanal (KK) and Mount Wilson (MW)
observatories. The data comprising eight and six solar cycles, respectively, were digitized by Howard et al. [\ref{Hea84},\ref{Hea99}] using the same technique
(ftp://ftp.ngdc.noaa.gov/STP/SOLAR\\ \_DATA/SUNSPOT\_REGIONS).
The digitizing technique [\ref{Hea84},\ref{Hea99}]
was aimed at using sunspots as tracers for measuring rotation and meridional flow.
Therefore, the digitized KK and MW catalogs consist
of pairs of datasets on sunspot groups observed on
two consecutive days. We wish to reconstruct the
evolution of the active regions from separate pairs of
datasets. Each pair of observations contains information
on a sunspot group observed on the first and
second day. If separate pairs of datasets correspond to
the same active region, the data for the second day of the preceding pair will coincide with the data for the first day of the suceeding pair. Accordingly, the evolution of the
sunspot groups was reconstructed by comparing the
dates of observations and the morphological features
of the active regions (group areas, numbers of spots,
group coordinates, etc) for the second day of the
preceding and the first day of following pairs of
datasets.

Using this method, however, the same sunspot group may be included
twice or even more times in the reconstructed data,
if the initial KK and MW data have gaps. On the other hand, if
a group was never observed during two successive
days, this group would be lost in the reconstructed
data. To assess possible multiple inclusions or losses
of some groups, we compared the total number of
sunspot groups in the reconstructed data with their
number in the catalog of the Royal Greenwich Observatory
(RGO) for the corresponding period of time (Table~1).
The initial MW and KK catalogs contain data
on sunspot groups located within $\pm 60^{\circ}$ of the central
meridian of the solar disk and observed on at least
two days. We applied the same restriction when
calculating the number of sunspot groups in the RGO
catalog. The differences in the numbers of sunspot
groups of about 200 for a total number of groups of about 15 000 are relatively small.

\bl {\small
\begin{description}
\item Table 1.
Total number of sunspot groups based on KK, MW, and GRO data for corresponding observation intervals
\end{description}
\centerline{
\begin{tabular}{|c|c|c|c|}
\hline Years & \multicolumn{3}{|c|}{Numbers of sunspot groups in catalogs}\\
\cline{2-4}
~~~~~~~~~&~~~~~~~RGO~~~~~~~~&~~~~~~~~~KK~~~~~~~~~&~~~~~~~~~MW~~~~~~~~~\\
\hline
1906--1987 &16195&15990&--\\
1917--1985 &14696&--&14552\\
\hline
\end{tabular}}}
\bl

The reconstructed data of the KK and MW catalogs
enabled us to follow the evolution of sunspot
groups and to estimate the sought for parameters of
the Babcock–Leighton mechanism for most of the
mature active regions. These catalogs also contain
the data on the spot numbers in groups, their areas,
coordinates, tilt angles, information on the
leading and following parts of the groups, etc.
The KK data cover the solar cycles 14-21, and MW data - cycles 16-21.
%%%%%%%%%%%%%%%%%%%%%%%%%%%%%%%%%%%%%%%%%%%%%%%%%%%%%%%%%%%%%%%%%%%%
 \bll
 \centerline{\sl 2.3. Results}
 \bl
%%%%%%%%%%%%%%%%%%%%%%%%%%%%%%%%%%%%%%%%%%%%%%%%%%%%%%%%%%%%%%%%%%%%

Figure~2 shows the positions of individual solar
cycles on a coordinate plane of the index $A$ of the large-scale field
for the solar minima succeeding these cycles versus
$B_{cyc}$ calculated with Eq.\,(\ref{1}). The $A$-index estimates
the amplitude of the large-scale magnetic field [\ref{MT00}]
(poloidal field for the epochs of solar minima).

\begin{figure}[htb]
 \centerline{
 \includegraphics[width=12cm]{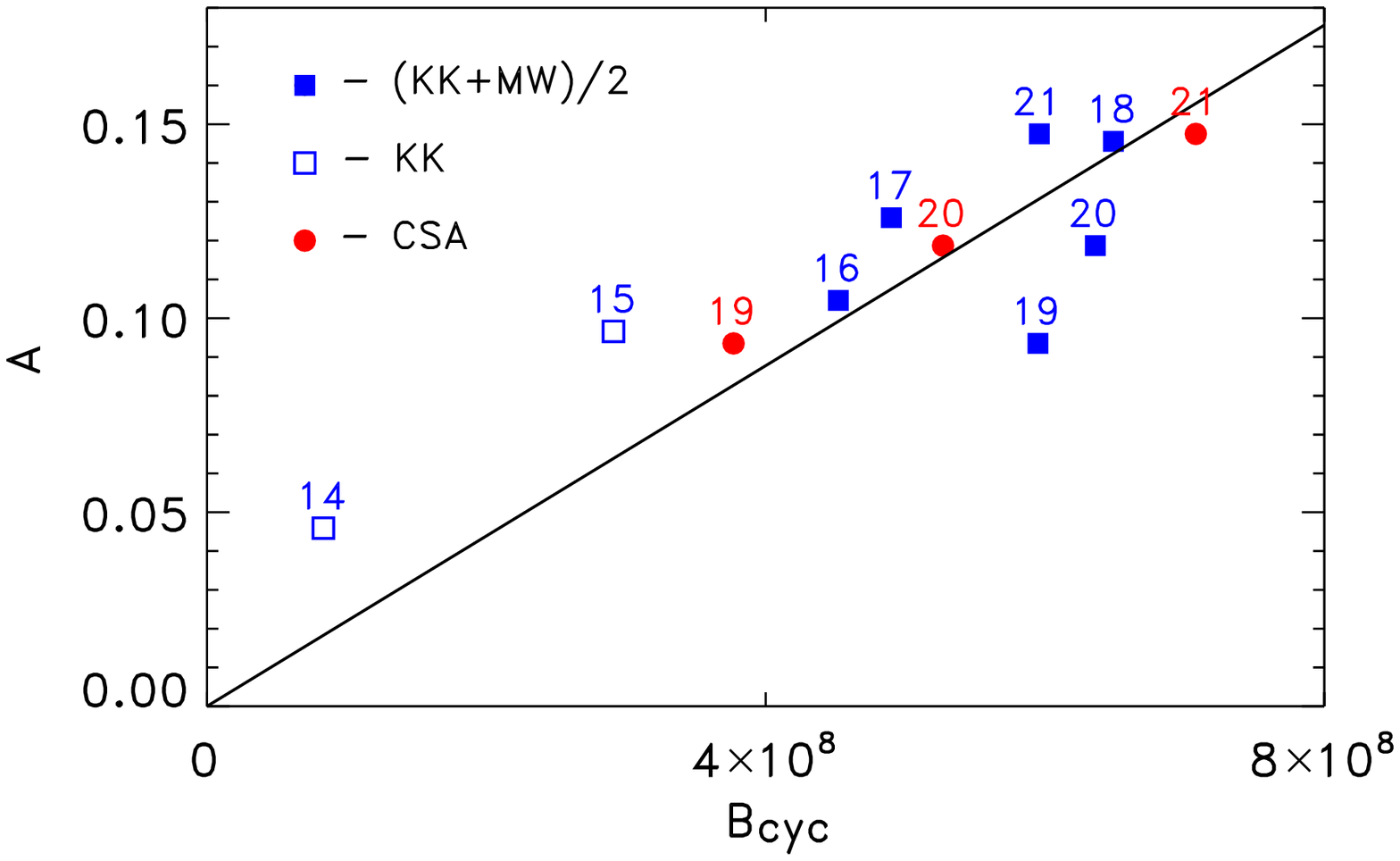}}
 \begin{description}
 \item{\small Fig.~2. Positions of individual solar cycles on the coordinate plane of $B_{cyc}$ and $A$-index of large-scale magnetic field for solar minima succeeding these cycles. The numbers near the points show the cycles No.
 }
 \end{description}
\end{figure}

The KK and MW catalogs give the areas of the
spot umbrae, while the CSA presents the total areas
including the penumbrae. Therefore, the CSA $B_{cyc}$
values in Fig.~2 have been reduced by a factor of 0.2.
A linear regression fit common for all the data,
\begin{equation}
    A = 2.16\times 10^{-10} B_\mathrm{cyc} ,
    \label{4}
\end{equation}
is also shown in Fig.~2. The KK and MW datasets are very similar. Common data points are therefore shown for solar cycles 16-21, for which KK and MW datasets overlap.  When calculating $B_{cyc}$ using
(\ref{1}), $\ell_i$ was taken in kilometers and $S_i$ in millionths
of a solar hemisphere. The coefficients $a$ of the linear fits, $A = a B_\mathrm{cyc}$,
for individual datasets, the correlation
coefficients and relative amplitude of the fluctuations
$\sigma_{B_\mathrm{Car}}$ (\ref{3}) for CSA data are given in Table~2.

\bl {\small
\begin{description}
\item Table~2. Correlation coefficients $R(B_{cyc},A)$, coefficients of linear regression
$a$, and standard deviation $\sigma_{B_\mathrm{Car}}$ (\ref{3}) calculated
for the KK, MW, and CSA data
\end{description}
\centerline{
\begin{tabular}{cccc}
\hline
Parameters&CSA&KK&MW\\
\hline
$R(B_{cyc},A)$    &0.98&0.81            &0.46\\
$a$               &$2.20 \times 10^{-10}$&$2.06 \times 10^{-10}$&$2.17 \times 10^{-10}$\\
$\sigma_{B_{car}}$&2.67&&\\
\hline
\end{tabular}}}
\bll

There is a well known correlation between the
index $A$ for solar minima and the amplitudes of the
succeeding solar cycles [\ref{MT00}-\ref{JCC07}]. The differential rotation
transforming the poloidal field into the toroidal field
varies only weakly with time, and does not contain
significant random fluctuations. Therefore, there is a functional
relation between the poloidal field at the solar minimum
and the toroidal field at the following maximum
(the sunspot activity is associated with the solar toroidal field). At the same time,
there is no functional relationship between the cycle amplitude
and the $A$-index of the following solar minimum [\ref{Mea01},\ref{C08}].
Randomness in the $\alpha$-effect providing such a relation may be the explanation for its lack of prominence.
This, however, does not preclude estimating the contribution of the $\alpha$-effect
to the poloidal field generation, including all the inherent fluctuations.
Formula (\ref{1}) provides such
an estimate for the special case of the a-effect named the Babcock–Leighton
mechanism. The estimated $B_{cyc}$ correlate well with the $A$-index.
Figure~2 and Table~2 suggest that the Babcock–Leighton mechanism is actually operating on the sun. Dasi-Espuig et al. [\ref{DEea10}] arrived at the same conclusion performing a quite different
analysis of KK and MW data.

\begin{figure}[htb]
 \centerline{
 \includegraphics[width=9cm]{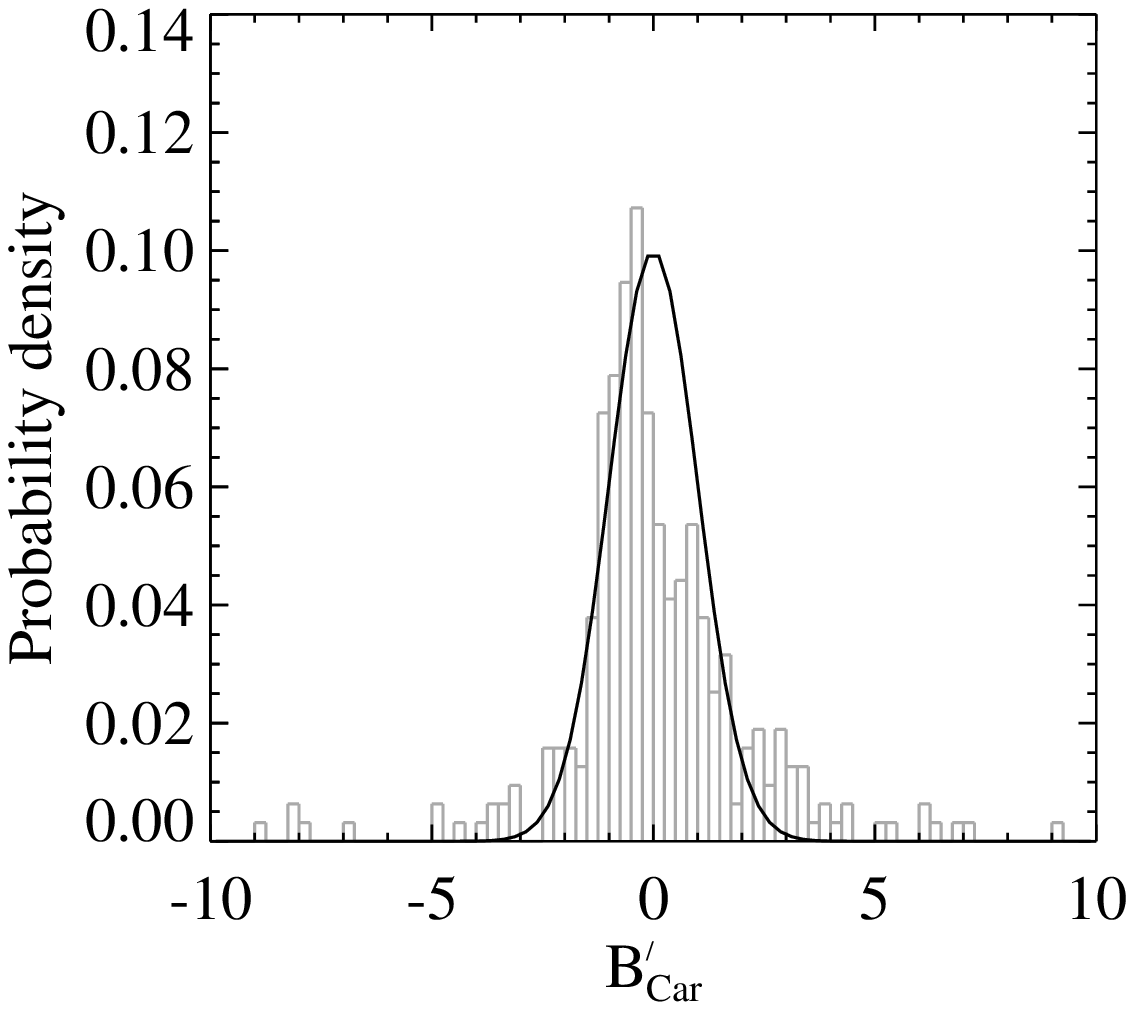}}
 \begin{description}
 \item{\small Fig.~3. Distribution of fluctuations $B'_\mathrm{Car}$ estimated from the CSA data.
    }
 \end{description}
\end{figure}

Let us now consider fluctuations of the $\alpha$-effect
for the Babcock–Leighton mechanism.
Figure~3 shows the distribution of the (relative)
fluctuations $B'_\mathrm{Car}$ (\ref{2}) estimated from the CSA
data. The distribution is close to normal. The relative amplitude (\ref{3}) of the fluctuation is given in Table\,3. We include fluctuations of the $\alpha$-effect with the estimated amplitude in the dynamo model described later.

Short-term fluctuations in solar-dynamo models are known to
result in alternations of durable epochs of comparatively
high and comparatively low amplitudes of magnetic
cycles. The fluctuations could be responsible for the stochastic
dynamics of the solar activity on time scales
of millennia [\ref{USK07}].

%%%%%%%%%%%%%%%%%%%%%%%%%%%%%%%%%%%%%%%%%%%%%%%%%%%%%%%%%%%%%%%%%%%%
 \bll
 \centerline{3. GRAND MINIMA AND MAXIMA IN THE DYNAMO MODEL}
%%%%%%%%%%%%%%%%%%%%%%%%%%%%%%%%%%%%%%%%%%%%%%%%%%%%%%%%%%%%%%%%%%%%
 \bl
 \centerline{\sl 3.1. The Model}
 \bl
%%%%%%%%%%%%%%%%%%%%%%%%%%%%%%%%%%%%%%%%%%%%%%%%%%%%%%%%%%%%%%%%%%%%
Our dynamo model is very close to that of an earlier paper [\ref{KO12}]. The only difference is that fluctuations of the $\alpha$-effect are now allowed for. The allowance for the fluctuations will, therefore, be discussed in detail, while other features of the model are outlined briefly. All further details can be found in [\ref{KO12}].

Our numerical model evolves the large-scale (longitude-averaged) magnetic
field with time in the spherical shell of the convective zone. We
assume axial symmetry of the field
 \begin{equation}
   {\vec B} = {\vec e}_\phi B + \rot \left({\vec
   e}_\phi\frac{\cal A}{r\sin\theta}\right) ,
   \label{5}
 \end{equation}
where $r,\theta$, and $\phi$ are the usual spherical coordinates, ${\vec
e}_\phi$ is the azimuthal unit vector, $B$ is the toroidal field, and
${\cal A}$ is the poloidal field potential. A similar expression for the fluid velocity,
 \begin{equation}
    \vec{V} = \vec{e}_\phi r\sin\theta\ \Omega f(r,\theta) +
    \frac{1}{\rho}\mathrm{rot}\left( \vec{e}_\phi
    \frac{\psi}{r\sin\theta}\right) ,
    \label{6}
 \end{equation}
accounts for rotation and meridional circulation. In this equation, $\Omega$ is the mean angular velocity, $f$ is the dimensionless rotational frequency, and $\psi$ is the stream
function of the meridional flow. The differential rotation is prescribed in accordance with  helioseismology and the meridional flow is specified in accordance with the model [\ref{KO11d}] for the global solar circulation.

Two specific properties of the model are the diamagnetic
transport of the field with the effective velocity
 \begin{equation}
    {\vec U}_\mathrm{dia} =
    -\frac{1}{2}{\vec\nabla}\eta_{_\mathrm{T}}
    \label{7}
 \end{equation}
[\ref{KR84}], where $\eta_{_\mathrm{T}}$ is the turbulent magnetic diffusivity, and
non-local formulation of the $\alpha$-effect of the
Babcock–Leighton type.

The dynamo equations are normalized to dimensionless variables.
The equation for the toroidal field,
 \begin{eqnarray}
    \frac{\partial B}{\partial t} &=&
    \frac{\eta}{x^2}\frac{\partial}{\partial\theta}\left(
    \frac{1}{\sin\theta}\frac{\partial(\sin\theta
    B)}{\partial\theta}\right)
    + \frac{1}{x}\frac{\partial}{\partial
    x}\left(\sqrt{\eta}\ \frac{\partial(\sqrt{\eta}\ xB)}
    {\partial x}\right) +
    \nonumber \\
    &+& \frac{R_\mathrm{m}}{x}\frac{\partial}{\partial\theta}
    \left(\frac{B}{\rho x\sin\theta}
    \frac{\partial\psi}{\partial x}\right)
    - \frac{R_\mathrm{m}}{x}\frac{\partial}{\partial x}
    \left(\frac{B}{\rho x\sin\theta}\frac{\partial\psi}{\partial\theta}
    \right) +
    \nonumber \\
    &+& \frac{\cal D}{x}
    \left(\frac{\partial f}{\partial x}\frac{\partial
    {\cal A}}{\partial\theta} - \frac{\partial f}{\partial\theta}
    \frac{\partial {\cal A}}{\partial x}\right) ,
 \label{8}
 \end{eqnarray}
includes two governing parameters: the dynamo number,
 \begin{equation}
    {\cal D} = \frac{\alpha_0 \Omega R_\odot^3}{\eta_0^2}\ -
 \label{9}
 \end{equation}
and the magnetic Reynolds number for the meridional flow,
\begin{equation}
    R_\mathrm{m} = \frac{V_0 R_\odot}{\eta_0}.
    \label{10}
\end{equation}
Here, $\alpha_0$ is the characteristic value of the $\alpha$-effect,
$\eta_0$ is the coefficient of turbulent diffusion at the
middle of the convection zone, and $V_0$ is the amplitude
of the meridional velocity. The time $t$
is measured in units of the diffusion time, $R^2_\odot /\eta_0$, and
$x = r/R_\odot$ is the relative radius.

All the computations were performed with  the
dynamo number ${\cal D} = 4.2\times 10^4$, which slightly
exceeds the critical value ${\cal D}_\mathrm{cr} = 3.96\times 10^4$ for which
the dynamo effect sets on. The Reynolds number
$R_\mathrm{m} = 10$, which corresponds to the amplitude of
the meridional flow $V_0 \simeq 14$\,m/s, assuming $\eta_0 \simeq 10^9$\,m$^2$/s.

The equation for the poloidal field,
 \begin{eqnarray}
    \frac{\partial {\cal A}}{\partial t} &=&
    \frac{\eta}{x^2}\sin\theta\frac{\partial}{\partial\theta}
    \left(\frac{1}{\sin\theta}\frac{\partial
    {\cal A}}{\partial\theta}\right) + \sqrt{\eta}\frac{\partial}{\partial
    x} \left(\sqrt{\eta}\frac{\partial {\cal A}}{\partial x}\right) +
    \nonumber \\
    &+& \frac{R_\mathrm{m}}{\rho x^2 \sin\theta}
    \left(\frac{\partial\psi}{\partial x}
    \frac{\partial {\cal A}}{\partial\theta} -
    \frac{\partial\psi}{\partial\theta}
    \frac{\partial {\cal A}}{\partial x}\right) +
    \nonumber \\
    &+& (1+s\sigma_{B_\mathrm{Car}}) x \sin^3\theta
    \cos\theta \int\limits_{x_\mathrm{i}}^x
    \alpha (x,x') B(x',\theta)\ \mathrm{d} x'  ,
 \label{11}
 \end{eqnarray}
differs from that used earlier in [\ref{KO12}] only by the presence
of finite $s\sigma_{B_\mathrm{Car}}$ in its last term. The finite $s\sigma_{B_\mathrm{Car}}$ takes into account the fluctuations of the $\alpha$-effect. In Eq.\,(\ref{11}),
$\sigma_{B_\mathrm{Car}}$ is the relative magnitude of the fluctuations (\ref{3})
and  $s$ is a random number with Gaussian distribution and rms value equal to one. The
normal distribution for $s$ was realized using the Box–Muller transformation [\ref{Pea92}]. The quantity $s$ remains constant
within the (dimensionless) time interval $\tau = 5\times 10^{-5}$,
which approximately corresponds to the period
of solar rotation. After time $\tau$ passed, $s$ takes
a new random value independent of its
preceding value. This new value remains constant
during time $\tau$, then is replaced by a new random
value for a time $\tau$, and so on. Thus, the fluctuations of the $\alpha$-effect
are modeled by a Poisson-type random process. This, however, does not mean that the variations of amplitudes of the magnetic cycles also obey a Poisson
distribution. The magnetic field varies continuously
with time, and the dynamo memory-time significantly
exceeds the correlation time for the random fluctuations
in the $\alpha$-effect. The relative magnitude of
the fluctuations of the $\alpha$-effect in the model is
inferred from the CSA data, $\sigma_{B_\mathrm{Car}} = 2.67$ (see
Table~2).

The function $\alpha (x,x')$ in (\ref{11}) characterizes the
non-local properties of the $\alpha$-effect. Similar to [\ref{KO12}],
this function was taken in the form
 \begin{equation}
    \alpha (x,x') = \frac{\phi_\mathrm{b} (x')\phi_\alpha (x)}{1 +
    B^2(x',\theta)},
    \label{12}
 \end{equation}
where $B^2$ in the denominator accounts for the
non-linear suppression of the $\alpha$-effect, the function
$\phi_\mathrm{b} (x')$ defines a spherical layer near the
base of the convective zone, whose toroidal field
produces the $\alpha$-effect, and the
function $\phi_\alpha (x)$ defines the subsurface region in which
this effect is produced. Figure~4 presents the functions $\phi_\mathrm{b}$
and $\phi_\alpha (x)$, together with the profile of the (normalized)
magnetic diffusion $\eta$ used in the dynamo model.

\begin{figure}[htb]
 \centerline{
 \includegraphics[width=10cm]{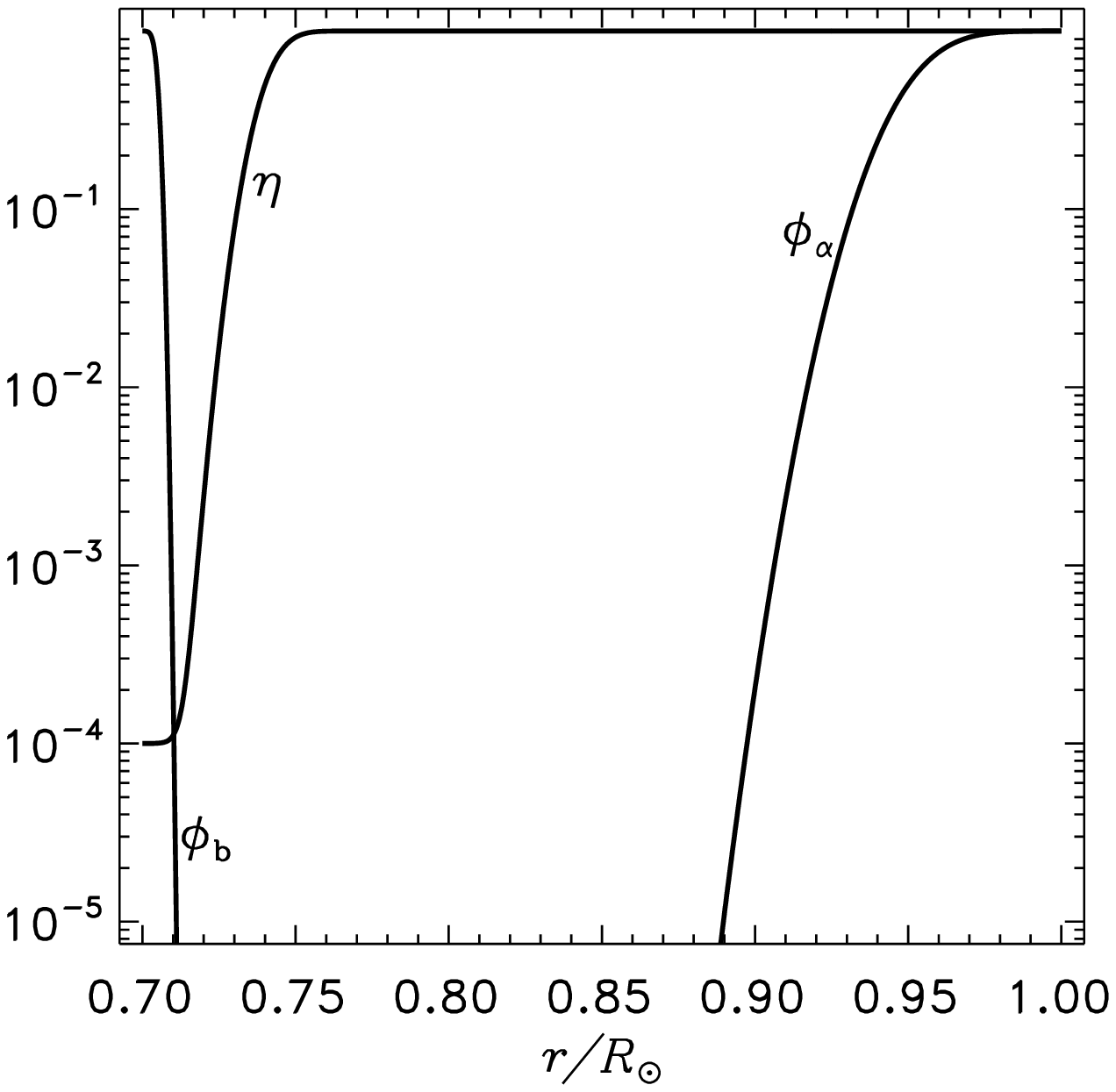}}
 \begin{description}
 \item{\small Fig.~4. The functions $\phi_\mathrm{b}$ and $\phi_\alpha$ and the profile of the magnetic diffusion $\eta$.
    }
 \end{description}
\end{figure}

The lower boundary is located at $x_\mathrm{i} = 0.7$. The
boundary conditions there correspond to the interface with a superconductor,
while the vacuum conditions were imposed on the top boundary.
We solved the dynamo equations (\ref{8}) and (\ref{11}) numerically applying
an explicit finite-difference scheme.

Taking $\sigma_{B_\mathrm{Cor}} = 0$ in (\ref{11}) eliminates fluctuations
of the $\alpha$-effect, and the dynamo model becomes identical to that
discussed in [\ref{KO12}]. This model can reproduce the main features of the solar cycle
fairly well. Here, we simulate global solar minima (and maxima)
by including fluctuations in the Babcock–Leighton $\alpha$-effect.

It is believed that solar active regions emerge
when deep toroidal fields rise to the solar surface.
Within each cycle, then, the number of sunspots varies in
phase with the the toroidal field. Since the
Babcock–Leighton mechanism is related to sunspot
activity, we assume that the number of sunspots
is proportional to the same magnetic flux of
the toroidal field, which defines the
$\alpha$-effect of the equations (\ref{11}) and (\ref{12}):
\begin{equation}
    B_{_\mathrm{W}} =
    \int\limits_{x_\mathrm{i}}^1\!\!\!\int\limits_0^\pi\! \sin\theta\,x\,\phi_\mathrm{b}(x)\,|B(x,\theta)|\,\mathrm{d}x\,\mathrm{d}\theta.
    \label{13}
\end{equation}
Here, the integrand in (\ref{13}) contains the absolute
value of $B(x,\theta)$, since the Wolf number is defined independent
of sunspot polarity. The relationship between the
Wolf number and $B_{_\mathrm{W}}$ (\ref{13}) is [\ref{Pip12}]
\begin{equation}
    W = C_{_W}\,B_{_W}\,exp(-{\frac{B_0}{B_{_W}}}).
    \label{14}
\end{equation}
The parameter values $C_{_W} = 10^5$ and $B_0 = 2 \times 10^{-5}$
provide the best agreement with the maximum and mean Wolf numbers
reconstructed from the radiocarbon $^{14}$C content in natural archives
[\ref{Sea04}]
(ftp://ftp.ncdc.noaa.gov/pub/data/paleo/climate\_forcing/solar \_variability).

%%%%%%%%%%%%%%%%%%%%%%%%%%%%%%%%%%%%%%%%%%%%%%%%%%%%%%%%%%%%%%%%%%%
 \bll
 \centerline{\sl 3.2. Grand Minima and Maxima}
 \bl
%%%%%%%%%%%%%%%%%%%%%%%%%%%%%%%%%%%%%%%%%%%%%%%%%%%%%%%%%%%%%%%%%%%
We computed the field evolution over a long-time
interval, encompassing approximately 11 000 years in
physical (dimensional) time. The model takes into account
fluctuations of the  $\alpha$-effect as described
above. These fluctuations result in variations of the
durations of the magnetic cycles, from about 7.3 to 15.1 years.
The calculated amplitudes of the cycles also vary.
Figure 5a shows the calculated magnetic flux (\ref{13})
as a function of time. The narrow peaks in Fig. 5a
correspond to individual magnetic cycles. We can see
that epochs of higher magnetic activity alternate with
epochs of weak magnetic fields.

\begin{figure}[htb]
 \centerline{
    \includegraphics[width=14cm]{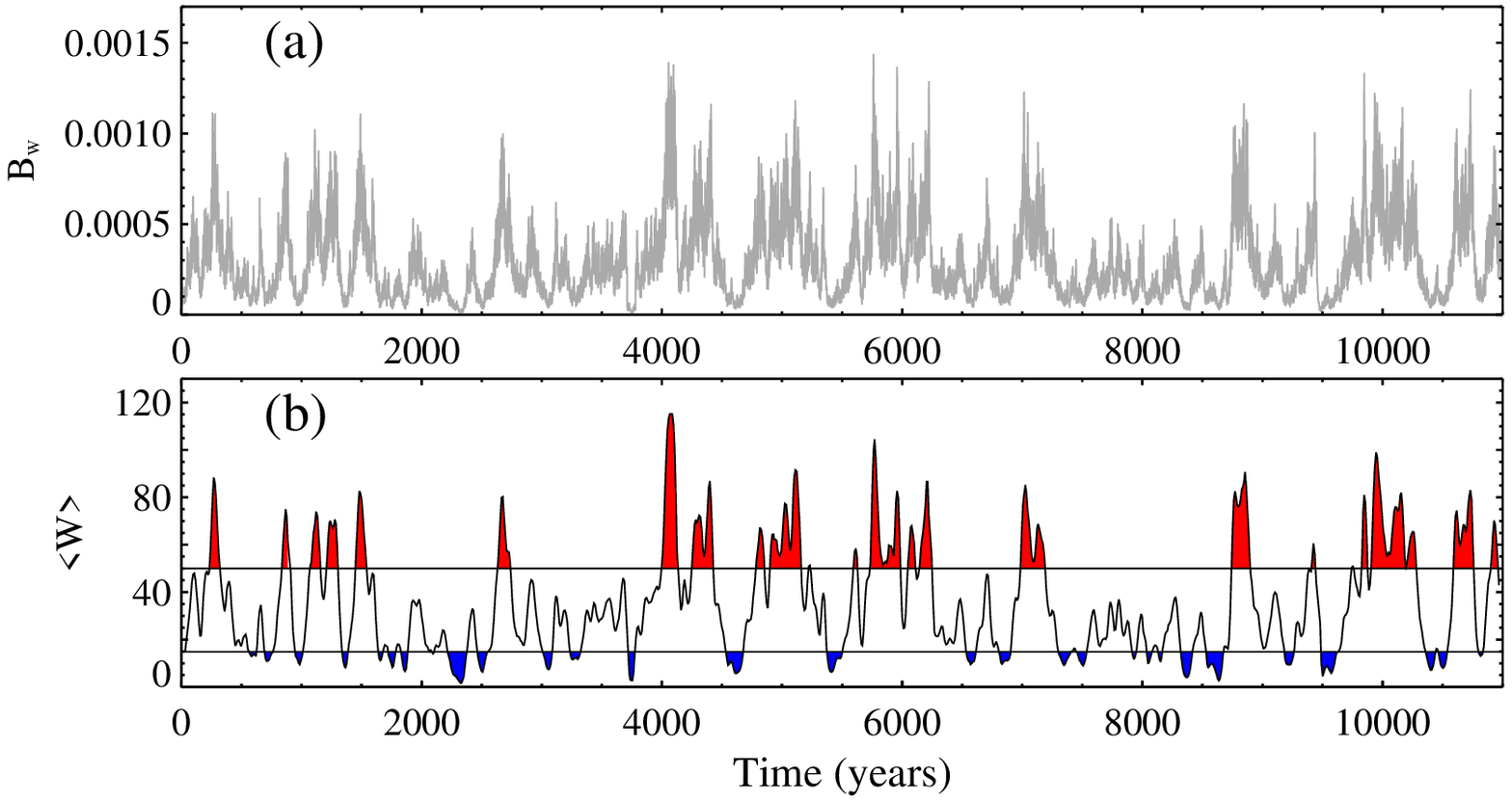}}
    \begin{description}
    \item{\small Fig.~5. The (a) magnetic flux $B_{_\mathrm{W}}(t)$ and (b) smoothed amplitudes $\langle W\rangle$ of the activity cycles as a function of time.
    }
 \end{description}
\end{figure}

To analyze the global minima and maxima of the dynamo model, the computed
function $B_{_\mathrm{W}}(t)$ was subject to a transformation following as
closely as possible the proces\-si\-ng applied to data on solar activity in
the remote past [\ref{USK07}]. We first convert $B_{_\mathrm{W}}$
into sunspot numbers by Eq.\,(\ref{14}). We then smooth the so defined "Wolf numbers"\ by computing its running mean over 13 solar rotations, i.e., over approximately one year. In analyses of long-term
variations of solar activity, the cycle amplitude is
usually defined as the maximum annual mean number of sunspots.
Various smoothing techniques were applied formerly to reveal secular and
super-secular variations, with Gleisberg secular smoothing
[\ref{VKK86}] being used most frequently. Part (b) of Fig.~5 shows the cycle
amplitude smoothed in this way as a function of time.
The rules for identifying the grand minima and maxima
are similar to those applied in [\ref{USK07}] to solar data. An epoch is identified as a grand minimum if $W$ was below
15 during at least two successive cycles ($> 20$ yrs). If the interval
between two neighboring minima was less than 30 yrs, such low-activity epochs
were combined into a single global minimum. In turn, global maxima were
defined as the epochs of $W$ exceeding 50. Global minima and maxima are indicated by
blue and read colours, respectively, in Fig.~5b and subsequent figures.

Figure~6a shows a histogram of the smoothed amplitudes of the
magnetic cycles $W$ (\ref{14}). 22\% of the calculated cycles have amplitudes $< 15$
to belong to grand minima, while $25\%$ with amplitudes $>50$ belong to grand maxima.

\begin{figure}[htb]
 \centerline{
 \includegraphics[width=6.5cm, height=6.0cm]{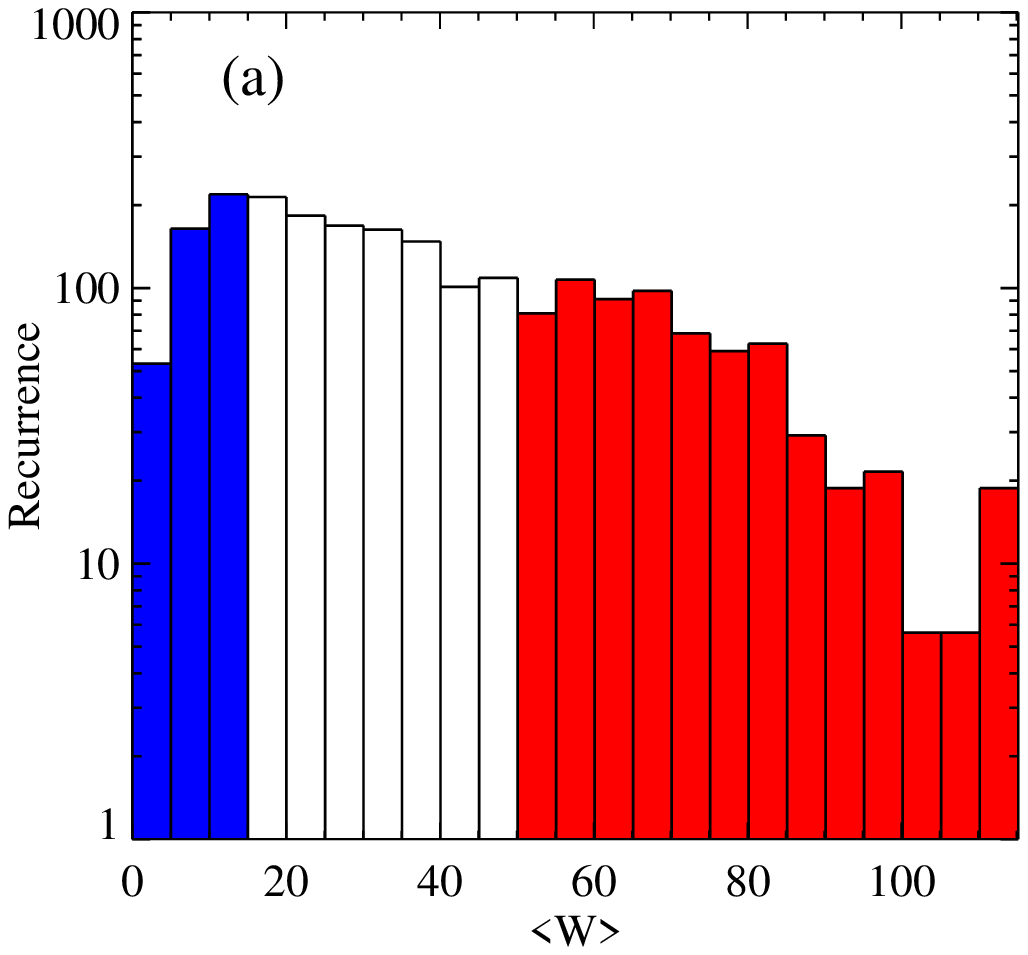}
 \hspace{0.5truecm}
 \includegraphics[width=6.5cm, height=6.0cm]{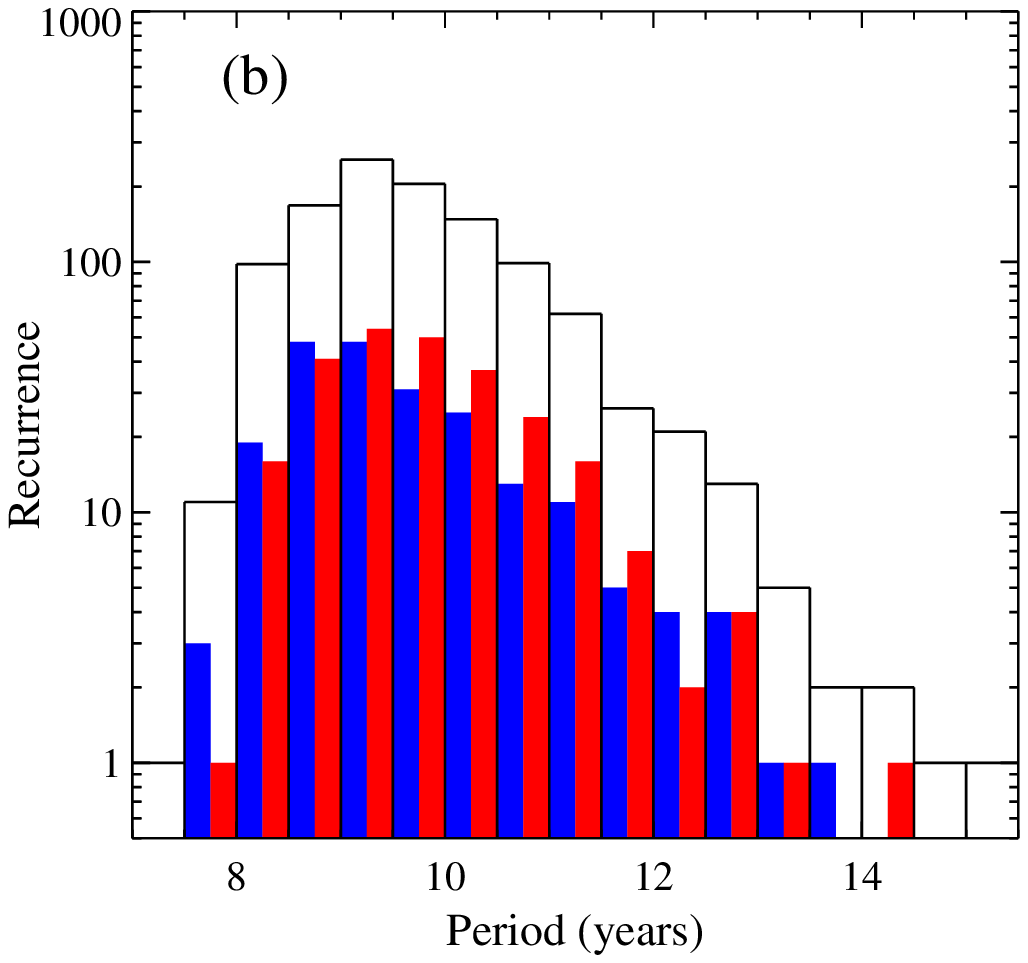}
 }
 \begin{description}
 \item{\small Fig.~6. Histograms of the (a) amplitudes and (b) durations of magnetic cycles. Blue and red correspond to the grand minima and grand maxima, respectively. Uncoloured histogram in part (b) shows total number of cycles disregarding their amplitudes.
    }
 \end{description}
\end{figure}

Figure~6b shows a histogram of durations of magnetic cycles.
The mean cycle duration in our model is about ten years,
somewhat shorter than the observed 11 years. The durations of the simulated
cycles vary from 7.3 to 15.1 yrs, but $93\%$ of the cycles lie
within the range of 8–11.5 yrs. Shorter cycles dominate during epochs of grand minima.
According to Nesme-Ribes et al. [\ref{NRea94}], durations of the cycles that emerged at
the end of the Maunder minimum were about 9–10 yrs. However, the
variations in solar activity deduced from the content of cosmogenic
isotopes in natural archives do not confirm this [\ref{Nea12}].

\begin{figure}[htb]
 \centerline{
 \includegraphics[width=7.0cm, height=6.0cm]{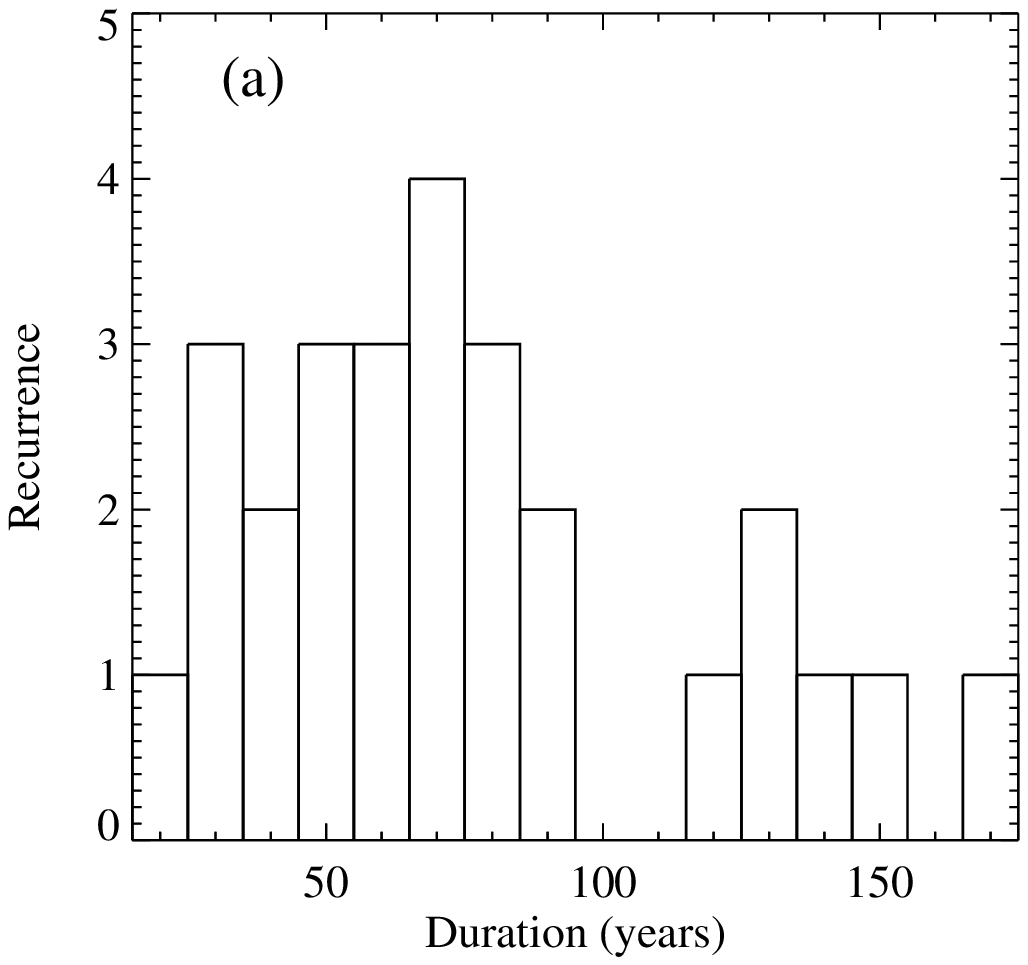}
 \hspace{0.5truecm}
 \includegraphics[width=7.0cm, height=6.0cm]{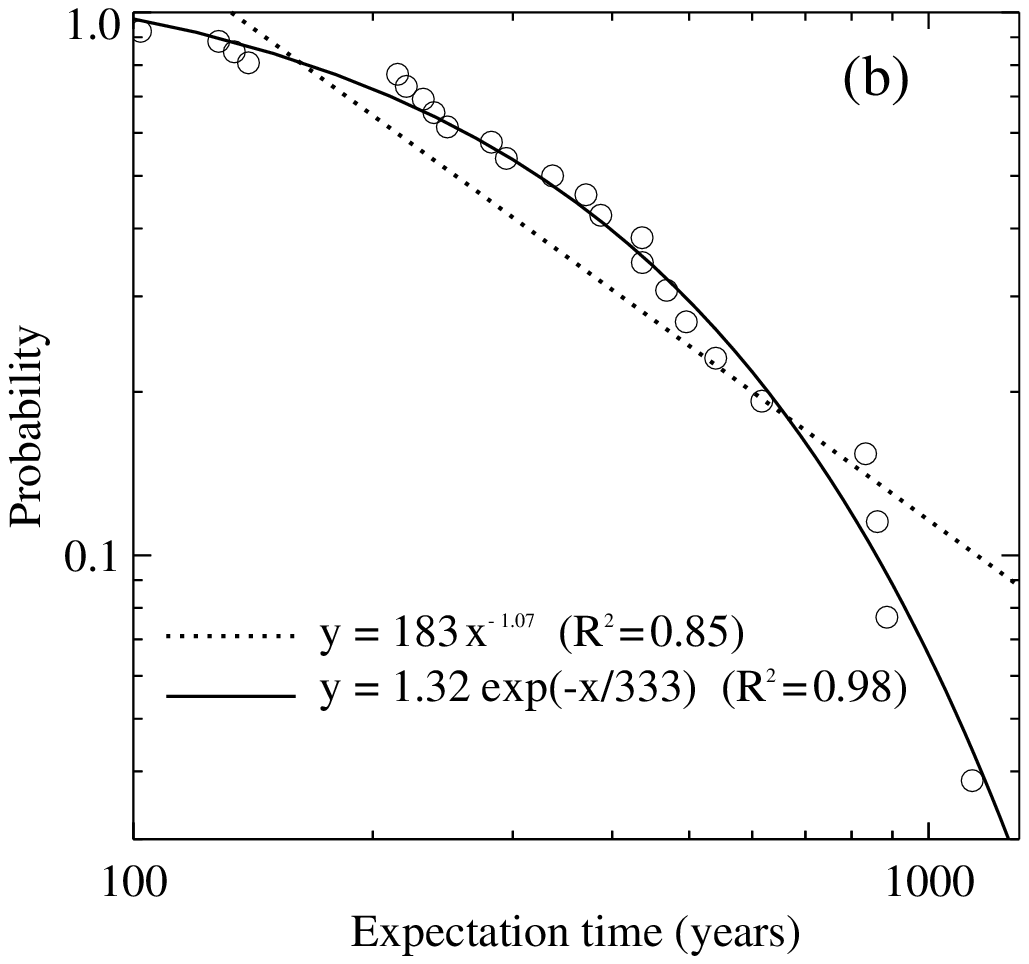}
 }
 \begin{description}
 \item{\small Fig.~7. (a) Histogram of the durations of the grand minima and (b)
   probability for the waiting time to be equal or longer than $t$ as function of $t$.
    }
 \end{description}
\end{figure}

Figure~7a shows the distribution of grand minima
durations. We can find relatively short minima
of $30-90$ yrs ($77.8\%$), as well as longer minima of
$> 110$ yrs ($22.2\%$). Similar groups of grand minima
were found from the cosmogenic isotopes data
[\ref{USK07},\ref{SB89}] and named the Maunder and Sperer type minima, respectively. In our computations, the total
duration of the grand minima equals 2009 yrs, reaching
$18.3\%$ of the computation time. The computations show 27 grand minima.

\begin{figure}[htb]
 \centerline{
 \includegraphics[width=7.0cm, height=6.0cm]{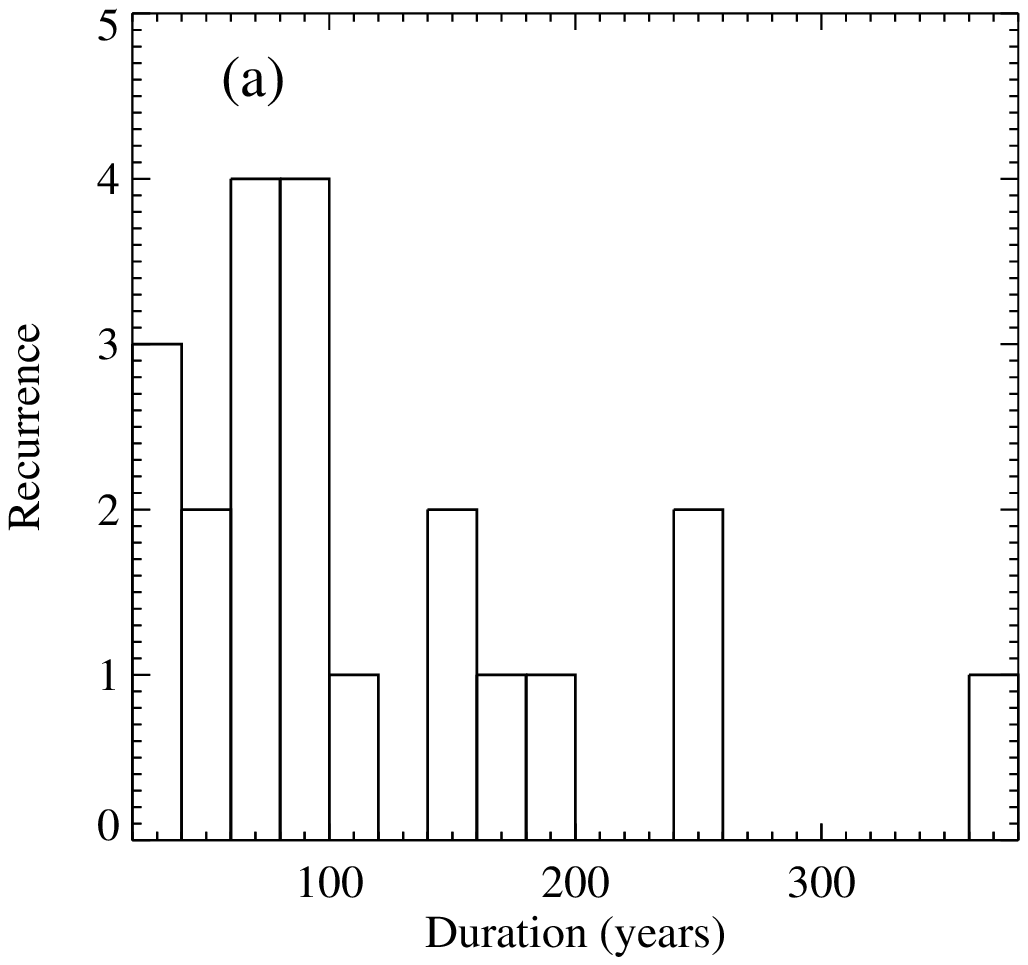}
 \hspace{0.5truecm}
 \includegraphics[width=7.0cm, height=6.0cm]{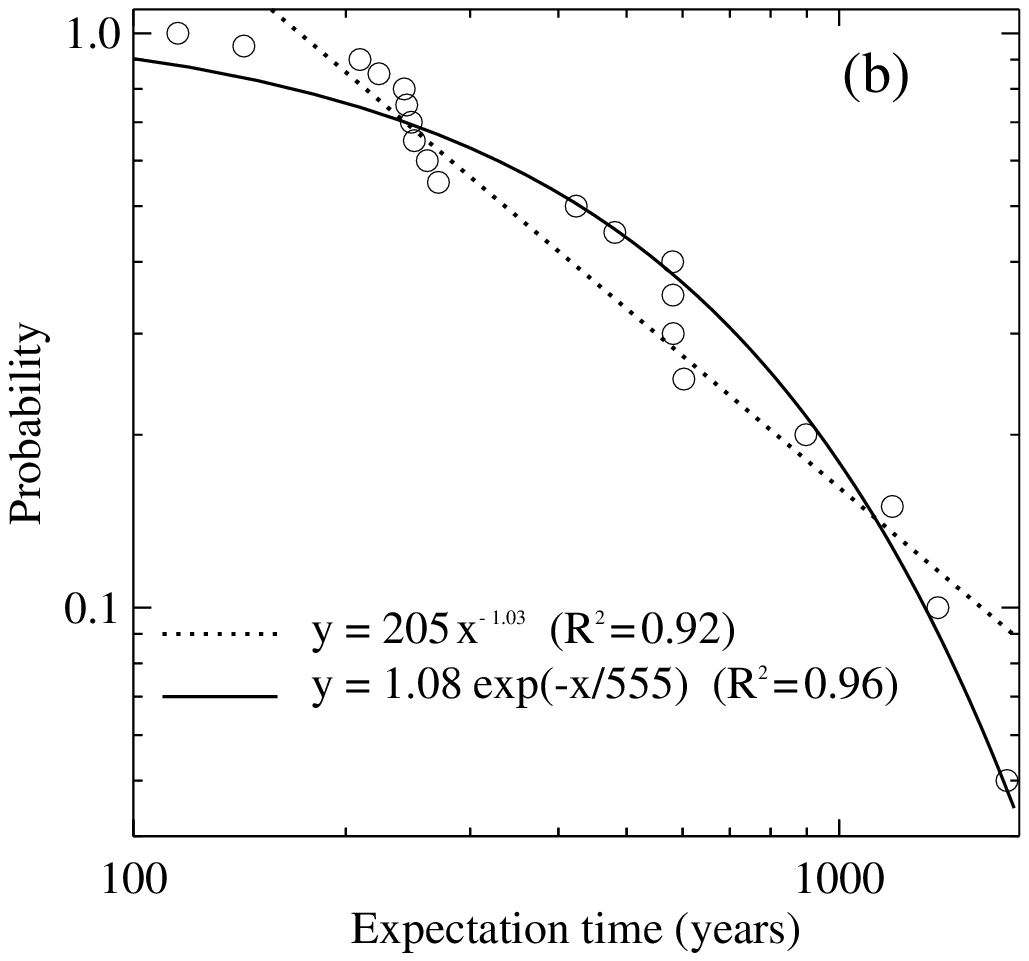}
 }
 \begin{description}
 \item{\small Fig.~8. Same as in Fig.~7 but for the grand maxima.
    }
 \end{description}
\end{figure}

Another important feature of global minima (maxima)
is their separation in time, called the waiting
time [\ref{USK07}]. The waiting time is defined as the time
interval $t$ between the centers of successive global
minima or maxima. Figures~7b and 8b show the
probabilities for the waiting time to be equal to or
longer than $t$ as a function of this $t$. For a finite number of observed or
simulated events, this probability can be estimated as
\begin{equation}
    y(t) = {\frac{N (t)}{N (0)}},
    \label{15}
\end{equation}
where $N(t)$ is the number of events with waiting
times not smaller than $t$. We approximated
the estimated probability distributions by exponential and
power-law functions:
\begin{equation}
    y(t)\propto \mathrm{exp}\left({\frac{-t}{T}}\right)
    \label{16}
\end{equation}
or
\begin{equation}
    y(t)\propto t^{-\gamma} .
    \label{17}
\end{equation}
The best-fit parameters we defined by the least-squares
method. The exponential distri\-bu\-t\-ion (\ref{16}) corresponds
to the Poisson random process, in which each successive
event occurs independently of preceding events. Such
a process has no memory of preceding events.

Figure~7b shows the distribution of the waiting
times for the grand minima. The parameter of the
exponential fit $T = 333\pm14$~yrs approximately corresponds
to the mean waiting time. The exponential
function describes the probability distribution better
than the power law.

Figure~8 shows a histogram of the durations and
the distribution of the waiting times for the grand
maxima. Similar to the case of the grand minima, the
distribution is close to a Poisson random process.
Analysis of the solar activity in the remote past leads
to a similar conclusion [\ref{USK07}]. Table~3 compares the
observed and computed results in more detail.

\bl {\small
\begin{description}
\item Table~3. Model computations compared with the solar activity data for about 11 000 years [\ref{USK07}]
\end{description}
\centerline{
\begin{tabular}{|c|c|c|}
\hline
Parameters                  &Model calculations   &Solar data ($^{14}$C)\\
\hline
{\bf Grand minima}         &                    &                      \\
Number                     &27                  &27                    \\
Total duration             &2009 years          &1880 years            \\
                           &$18\%$              &$17\%$                \\
Mean duration              &74 years            &70 years              \\
Waiting time               &$T=333\pm14$ years  &$T=435\pm15$ years    \\
                           &$\Gamma=1.07\pm0.09$&$\Gamma=0.95\pm0.02$  \\
\hline
{\bf Grand maxima}         &                    &                      \\
Number                     &21                  &19, 22$^*$            \\
Total duration             &2484 years          &1030 years, 1560 years$^*$\\
                           &$23\%$              &$9\%,\ 22\%^*$        \\
Waiting time               &$T=555\pm24$ years    &$T=355\pm20$ years      \\
                           &$\Gamma=1.03\pm0.07$&$\Gamma=0.77\pm0.05$  \\
\hline
\end{tabular}}
~~~~~$^*$ Calculated using 7000 years long data series.} \bll

Though the fluctuations of the $\alpha$-effect in our
model represent a Poisson-type random process, the nearly Poisson
distribution found for the grand minima (and maxima) is
not obvious. The magnetic field varies continuously
in the dynamo process and undoubtedly "remembers"\
its preceding states. Nevertheless, random
fluctuations in the $\alpha$-effect result in memory loss on
time scales exceeding the cycle period.

\begin{figure}[htb]
 \centerline{
 \includegraphics[width=13cm, height=9.0cm]{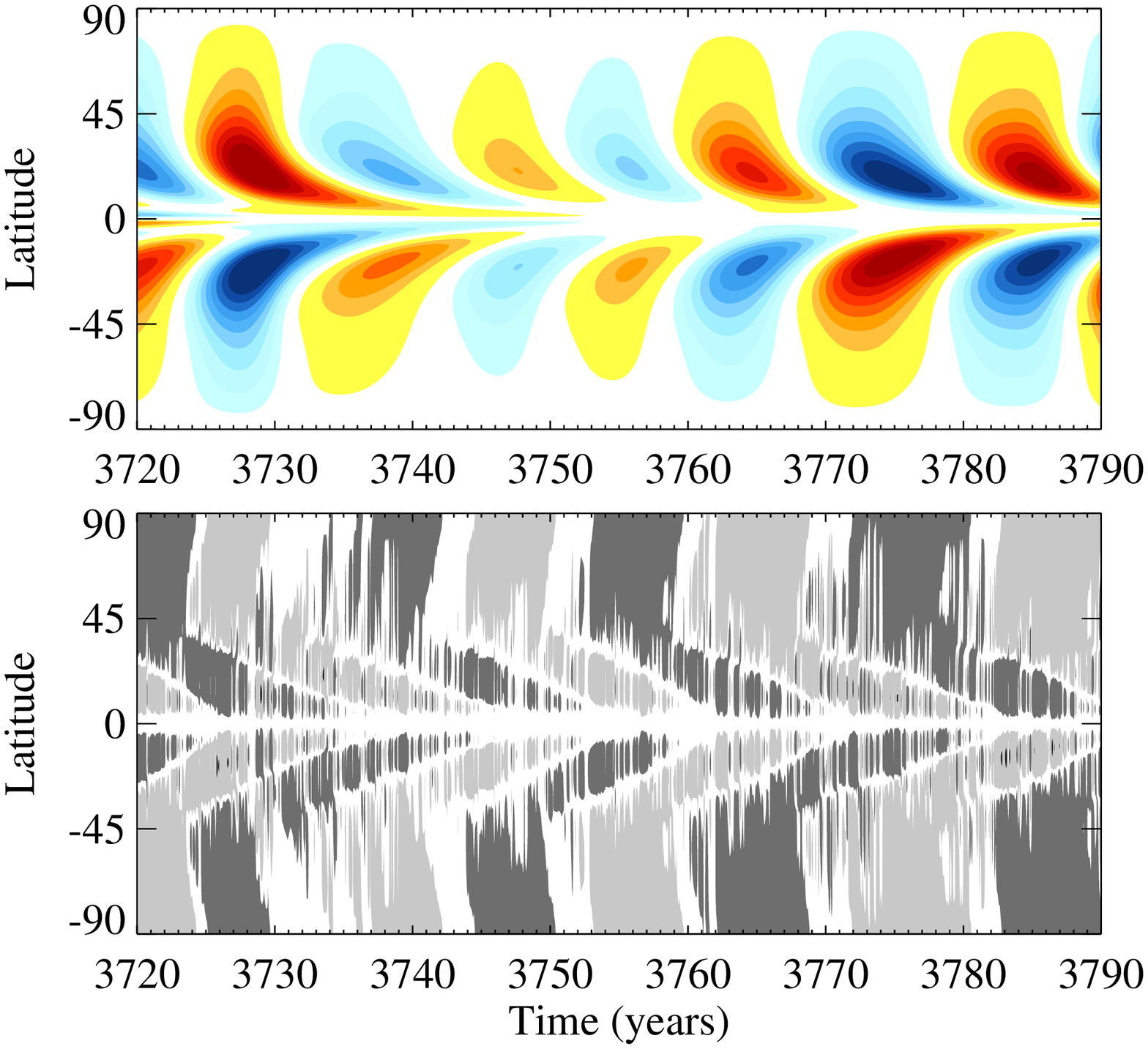}
 }
 \begin{description}
 \item{\small Fig.~9. Time-latitude diagrams for the toroidal field at the base of
   the convective zone (top panel) and the radial field at the surface (bottom).
   Yellow and read (blue) colours for toroidal field and dark (light) shading for the radial field correspond to positive (negative) sign of the fields.
    }
 \end{description}
\end{figure}

Figure~9 shows time-latitude diagrams for the toroidal field at the
base of the convective zone and for the radial field at the solar surface. The radial
(poloidal) field demonstrates both regular variations in the dynamo cycles and
irregular changes over shorter time scales comparable to the solar rotation
period. The poloidal field is generated by the $\alpha$-effect, and its irregular changes
are due to the random fluctuations in the $\alpha$-effect of our model. The toroidal field
is generated by the (steady) differential rotation. Therefore, the deep toroidal
field in Fig.~9 shows only almost periodic cycles with a slowly varying amplitude.
%%%%%%%%%%%%%%%%%%%%%%%%%%%%%%%%%%%%%%%%%%%%%%%%%%%%%%%%%%%%%%%%%%%%
 \bll
 \centerline{4. CONCLUSIONS}
 \bl
%%%%%%%%%%%%%%%%%%%%%%%%%%%%%%%%%%%%%%%%%%%%%%%%%%%%%%%%%%%%%%%%%%%%
The above estimates of the Babcock–Leighton mechanism based on data from three
sunspot catalogs support the idea that this mechanism operates on the Sun.
This mecha\-n\-ism for the generation of the poloidal field is a particular kind of
the $\alpha$-effect of hydromag\-ne\-tic dynamos. An important feature of the
Babcock–Leighton $\alpha$-effect is its non-local character: the generation
of the poloidal field near the solar surface is associated with the toroidal
field located at the base of the convective zone. The non-local $\alpha$-effect enables
us to improve the agreement between dynamo models and solar observations [\ref{KO12}].

Our calculations show that taking into account fluctuations of the (non-local) $\alpha$-effect
enables us to reproduce global changes in solar activity on time scales of centuries.
Here, we have used fluctuation parameters estimated from sunspot data. The fluctuations
in the Babcock–Leighton $\alpha$-effect are not small; their amplitude appreciably exceeds the mean
value. These fluctuations occur on comparatively short time scales of the order of
the solar rotation period. Our model calculations show that global
changes in solar activity similar to the Maunder minimum can be
caused by irregular variations in dynamo parameters on a time scale of the order of the
solar rotation period.

Our proposed model takes into account only the rough characteristics
of the fluctuations of the $\alpha$-effect. The model includes
temporal fluctuations and neglects irregular spatial changes, which
certainly are also present. Nevertheless, the
parameters of the grand solar minima (maxima) calculated are in overall agreement with
data on solar activity in the remote past (see Table~3). At the
same time, the model proposed cannot reproduce fine features
of global changes, such as the breaking of the equatorial symmetry in solar
activity as observed at the end of the Maunder minimum [\ref{SNR94},\ref{A09}].
This effect requires accounting for irregular changes in the
$\alpha$-effect with latitude, which can be a perspective
for development of the model.
%%%%%%%%%%%%%%%%%%%%%%%%%%%%%%%%%%%%%%%%%%%%%%%%%%%%%%%%%%%%%%%%%%%
\bll

This work was supported by the Russian Foundation for Basic Research
(projects 12-02-92691-IND\_a, 13-02-00277, and 12-02-33110-
mol\_a\_ved) and the Ministry of Education and Science of the
Russian Federation (state contract No 16.518.11.7065).
%%%%%%%%%%%%%%%%%%%%%%%%%%%%%%%%%%%%%%%%%%%%%%%%%%%%%%%%%%%%%%%%%%%%
 \bll
 \centerline{REFERENCES}
 \begin{enumerate}
 \item\label{KR84}
    F.~Krauze and K.-H.~R\"adler,
    {\it Mean Field Electrodynamics and Dynamo Theory} (Pergamon, Oxford,
    1980; Mir,Moscow, 1984).
 \item\label{LH82}
    B.\,J.~Labonte and R.~Howard,
    Sol. Phys. {\bf 75}, 161 (1982).
 \item\label{Vea02}
    S.\,V.~Vorontsov, J.~Christensen-Dalsgaard, J.~Schou, {\it et al.},
    Science {\bf 296}, 101 (2002).
 \item\label{MT00}
    V.I.~Makarov and A.G.~Tlatov,
    Astron. Rep. {\bf 44}, 759 (2000).
  \item\label{Mea01}
    V.\,I.~Makarov, A.\,G.~Tlatov, D.\,K.~Callebaut, {\it et al.},
    Sol. Phys. {\bf 198} 409 (2001).
 \item\label{JCC07}
    J.~Jiang, P.~Chatterjee, and A.\,R.~Choudhuri,
    MNRAS {\bf 381}, 1527 (2007).
 \item\label{C11}
    A.\,R.~Choudhuri,
    Proc. IAU Symp. 273 {\it The Physics of Sun and Star Spots}.
    \\ D.~Choudhary \& K.~Strassmeier (eds.) Kluwer, Dordrecht, p.28
    (2011).
 \item\label{P55}
    E.\,N.~Parker,
    Astrophys. J. {\bf 122}, 293 (1955).
 \item\label{B61}
    H.\,W.~Babcock,
    Astrophys. J. {\bf 133}, 572 (1961).
 \item\label{L69}
    R.\,B.~Leighton,
    Astrophys. J. {\bf 156}, 1 (1969).
 \item\label{KO11a}
    L.L.~Kitchatinov and S.V.~Olemskoy,
    Astron. Lett. {\bf 37}, 286 (2011).
 \item\label{KO11b}
    L.\,L.~Kitchatinov and S.\,V.~Olemskoy,
    Astron. Nachr. {\bf 332}, 496 (2011).
 \item\label{E04}
    D.\,V.~Erofeev,
    Proc. IAU Symposium 223 {\it Multi-Wavelength Investigations of Solar Activity}.
    A.\,V.~Stepanov, E.\,E.~Benevolenskaya \& A.\,G.~Kosovichev (eds.)
    Kluwer, Dordrecht, p.97 (2004).
 \item\label{DEea10}
    M.~Dasi-Espuig, S.\,K.~Solanki, N.\,A.~Krivova, {\it et al.},
    Astron. Astrophys. {\bf 518} A7 (2010).
 \item\label{KO11c}
    L.L.~Kitchatinov and S.V.~Olemskoy,
    Astron. Lett. {\bf 37}, 656 (2011).
 \item\label{SB92}
    S.\,H.~Saar and S.\,L.~Baliunas,
    {\it The Solar Cycle Workshop}. K.\,L.~Harvey (ed.) ASP Conf.
    Series {\bf 27}, 150 (1992).
 \item\label{C92}
    A.\,R.~Choudhuri,
    Astron. Astrophys. {\bf 253}, 277 (1992).
 \item\label{OHS96}
    A.\,J.\,H.~Ossendrijver, P.~Hoyng, and D.~Schmitt,
    Astron. Astrophys. {\bf 313}, 938 (1996).
 \item\label{Mea08}
    D.~Moss, D.~Sokoloff, I.~Usoskin, and V.~Tutubalin,
    Sol. Phys. {\bf 250}, 221 (2008).
 \item\label{USM09}
    I.\,G.~Usoskin, D.~Sokoloff, and D.~Moss,
    Sol. Phys. {\bf 254}, 345 (2009).
 \item\label{MBT06}
    M.\,S.~Miesch, A.\,S.~Brun, and J.~Toomre,
    Astrophys. J. {\bf 641}, 618 (2006).
 \item\label{WS89}
    Y.-M.~Wang and N.\,R.~Sheeley\,Jr.,
    \sp\ {\bf 124}, 81 (1989).
 \item\label{O85}
    V.N.~Obridko,
    {\sl Solar Spots and Activity Complexes} (Nauka, Moscow, 1985) [in Russian].
  \item\label{Nea08}
    Yu.A.~Nagovitsyn, E.V.~Miletskii, V.G.~Ivanov, and S.A.~Guseva,
    Kosmich. Issled. {\bf 46}, 291 (2008).
 \item\label{Hea84}
    R.~Howard, P.\,I.~Gilman, and P.\,A.~Gilman,
    \apj\ {\bf 283}, 373 (1984).
 \item\label{Hea99}
    R.\,F.~Howard, S.\,S.~Gupta, and K.\,R.~Sivaraman,
    \sp\ {\bf 186}, 25 (1999).
 \item\label{C08}
    A.\,R.~Choudhuri,
    J. Astrophys. Astr. {\bf 29}, 41 (2008).
 \item\label{USK07}
    I.\,G.~Usoskin, S.\,K.~Solanki, and G.\,A.~Kovaltsov,
    Astron. Astrophys. {\bf 471}, 301 (2007).
 \item\label{KO12}
    L.\,L.~Kitchatinov and S.\,V.~Olemskoy,
    Sol. Phys. {\bf 276} 3 (2012).
 \item\label{KO11d}
    L.\,L.~Kitchatinov and S.\,V.~Olemskoy,
    MNRAS {\bf 411} 1059 (2011).
 \item\label{Pea92}
    W.\,H.~Press, S.\,A.~Teukolsky, W.\,T.~Vetterling, and
    B.\,P.~Flannery,
    {\it Numerical Recipes}. Cambridge Univ. Press (1992).
  \item\label{Pip12}
    V.\,V.~Pipin, D.\,D.~Sokolov, and I.\,G.~Usoskin,
    \aa\ {\bf 542}, A26 (2012).
 \item\label{Sea04}
    S.\,K.~Solanki, I.\,G.~Usoskin, B.~Kromer, et. al.,
    \nat\ {\bf 431}, 1084 (2004).
 \item\label{VKK86}
    Yu.I.~Vitinskii, M.~Kopetskii, and G.V.~Kuklin,
    {\sl Statistics of Spot-Forming Activity of the Sun}
    (Nauka, Moscow, 1986) [in Russian].
 \item\label{NRea94}
    E.~Nesme-Ribes, D.~Sokoloff, J.\,C.~Ribes, and M.~Kremliovsky,
    in {\sl The Solar Engine and its Influence on Terrestrial Atmosphere and Climate.} NATO ASI
    Ser., Vol. I~25, E.~Nesme-Ribes (ed.) Springer,
    Berlin, p.71 (1994).
 \item\label{Nea12}
    K.~Nagaya, K.~Kitazawa, F.~Miyake, et al.,
    \sp\ {\bf 280}, 223 (2012).
 \item\label{SB89}
    M.~Stuiver and T.\,F.~Braziunas,
    \nat\ {\bf 338}, 405 (1989).
 \item\label{SNR94}
    D.~Sokoloff and E.~Nesme-Ribes,
    \aa\ {\bf 288}, 293 (1994).
 \item\label{A09}
    R.~Arlt,
    \sp\ {\bf 255}, 143 (2009).
\end{enumerate}
%%%%%%%%%%%%%%%%%%%%%%%%%%%%%%%%%%%%%%%%%%%%%%%%%%%%%%%%%%%%%%%%%%%

\end{document}